\RequirePackage{rotating}
\documentclass[twocolumn,useAMS,usenatbib, 12pt]{aastex6}
\usepackage{graphicx, subfigure}

\usepackage{lipsum}
\usepackage{array,booktabs,dcolumn,calc}
\usepackage[fleqn]{amsmath}
\usepackage{float}
\usepackage{bm}
\usepackage{amssymb,amsmath}
\usepackage{setspace}
\usepackage{rotating}
\usepackage{xcolor}
\usepackage{bold-extra}
\hypersetup{colorlinks = true, citecolor=black, linkcolor=blue}

\begin{document}

\title{Removing visual bias in filament identification: a new goodness-of-fit measure}

\author{C.-E. Green\altaffilmark{1,2}, M.~R. Cunningham\altaffilmark{1}, J.~R. Dawson\altaffilmark{2,3}, P.~A. Jones\altaffilmark{1}, G. Novak\altaffilmark{4}, L.~M. Fissel\altaffilmark{4}}

\altaffiltext{1}{School of Physics, University of New South Wales, Sydney, NSW, 2052, Australia. Email: claire.elise.green@gmail.com}
\altaffiltext{2}{CSIRO Astronomy \& Space Science, Australia Telescope National Facility, PO Box 76, Epping, NSW 1710, Australia}
\altaffiltext{3}{Department of Physics and Astronomy and MQ Research Centre in Astronomy, Astrophysics and Astrophotonics, Macquarie University, NSW 2109, Australia}
\altaffiltext{4}{Center for Interdisciplinary Exploration and Research in Astrophysics (CIERA) and Department\ of Physics \& Astronomy, Northwestern University, 2145 Sheridan Road, Evanston, IL 60208, U.S.A.}

\begin{abstract}
Different combinations of input parameters to filament identification algorithms, such as \textsc{disperse} and \textsc{filfinder}, produce numerous different output skeletons. The skeletons are a one pixel wide representation of the filamentary structure in the original input image. However, these output skeletons may not necessarily be a good representation of that structure. Furthermore, a given skeleton may not be as good a representation as another. Previously there has been no mathematical `goodness-of-fit' measure to compare output skeletons to the input image. Thus far this has been assessed visually, introducing visual bias. We propose the application of the mean structural similarity index (MSSIM) as a mathematical goodness-of-fit measure. We describe the use of the MSSIM to find the output skeletons most mathematically similar to the original input image (the optimum , or `best', skeletons) for a given algorithm, and independently of the algorithm. This measure makes possible systematic parameter studies, aimed at finding the subset of input parameter values returning optimum skeletons. It can also be applied to the output of non-skeleton based filament identification algorithms, such as the Hessian matrix method. The MSSIM removes the need to visually examine thousands of output skeletons, and eliminates the visual bias, subjectivity, and limited reproducibility inherent in that process, representing a major improvement on existing techniques. Importantly, it also allows further automation in the post-processing of output skeletons, which is crucial in this era of `big data'.

\end{abstract}

\section{Introduction} \label{sec:intro}
The advent of the Herschel\footnote{Herschel is an ESA space observatory with science instruments provided by European-led Principal Investigator consortia and with important participation from NASA.} Space Observatory \citep{pilbratt2010} has allowed detailed mapping of the structure of stellar nurseries, molecular clouds. Herschel observations have affirmed the filamentary nature of these clouds and suggested these structures may play an important role in star formation \citep[e.g.][]{molinari2010, arzoumanian2011, hill2011, palmeirim2013}. This has motivated a boom in the study of these filamentary structures in infra-red dust and radio molecular line data \citep[e.g.][]{cox2016, busquet2013}. At the foundation of the study of these thread-like structures lies our ability to automatically identify them in such data. The construction of a skeleton tracing the filamentary structures is facilitated by filament finding algorithms such as \textsc{disperse} \citep{sousbie2011} and \textsc{filfinder} \citep{koch2015}. The lack of a goodness-of-fit measure to apply to skeletons output by these algorithms represents a major problem in this field. \\

The filament skeleton is a one pixel wide representation of the filaments structure, tracing the main filament and its branches. When using the same input image (e.g. a dust column density map) different combinations of input parameters to filament finding algorithms produce different output skeletons. Different algorithms also return different skeletons. This is illustrated in~\autoref{fig:skel_examples} and~\autoref{fig:skel_examples_disperse} where we present the input image (Herschel dust column density map for the South-ridge of Vela C of \citet{fissel2016}), alongside the output skeletons produced by \textsc{filfinder} and \textsc{disperse} with different input parameter combinations. \\

The filament skeleton forms the foundation for other analyses. These include  measurements of length, width and mass per unit length \citep[e.g.][]{arzoumanian2011}, and relative orientation comparisons with respect to magnetic fields and clumps (\citealt{green2017}, in preparation). These analyses are then interpreted in terms of supporting or rejecting star formation hypotheses. It is therefore essential that the filament skeleton is as good a representation of the structures in the input image as possible. \\

Current works involving filament skeletons can be divided into two groups: 1) those that repeatedly perform an analysis (like width fitting) on multiple skeletons for the same input image, and 2) those that perform the analysis on a single skeleton. In general the former perform a parameter study\footnote{By parameter study we refer to the process of repeatedly running the filament finding algorithm with different combinations of input parameters, which sample the sensible values for those inputs.}, and visually identify the subset of skeletons that most reasonably reproduce the structures within the original input image \citep[e.g.][]{panopoulou2014}. Analyses are then performed on each skeleton in that selected subset. The latter may also perform a parameter study, and then either a single output skeleton corresponding to one set of input parameters is chosen visually as most closely reproducing the structures of the original image  \citep[e.g.][]{hill2011} or constraints on a property (such as filament/skeleton length) may be used to whittle the larger pool of output skeletons down to a smaller set, which can then be combined \citep[e.g.][]{liu2016}.\\

These approaches all involve an assessment of similarity between the output skeletons and the structures in the original input image. The previous lack of a mathematical goodness-of-fit measure forced astronomers to make this assessment visually, and to hence potentially introduce visual bias. We propose that the mean structural similarity index (MSSIM) can be used as a goodness-of-fit measure to find the mathematically most similar output skeleton(s) to the input image, solving this problem. The MSSIM will be discussed in~\autoref{sec:SSIM}. This measure makes it possible to examine the relative similarities of the output skeletons (with respect to the input image) as a function of the input parameters, in order
to find the range of optimum input values in parameter studies, as will be demonstrated in~\autoref{sec:optimisation}.  \\


\section{Mathematical similarity with the MSSIM} \label{sec:SSIM}
\subsection{Current filament finding algorithms} \label{sec:current_filfinding}

\begin{figure*}[p]
       \centering
       
      \includegraphics[width=\textwidth, clip=true, trim=1.8cm 4.5cm 1.8cm 4cm]{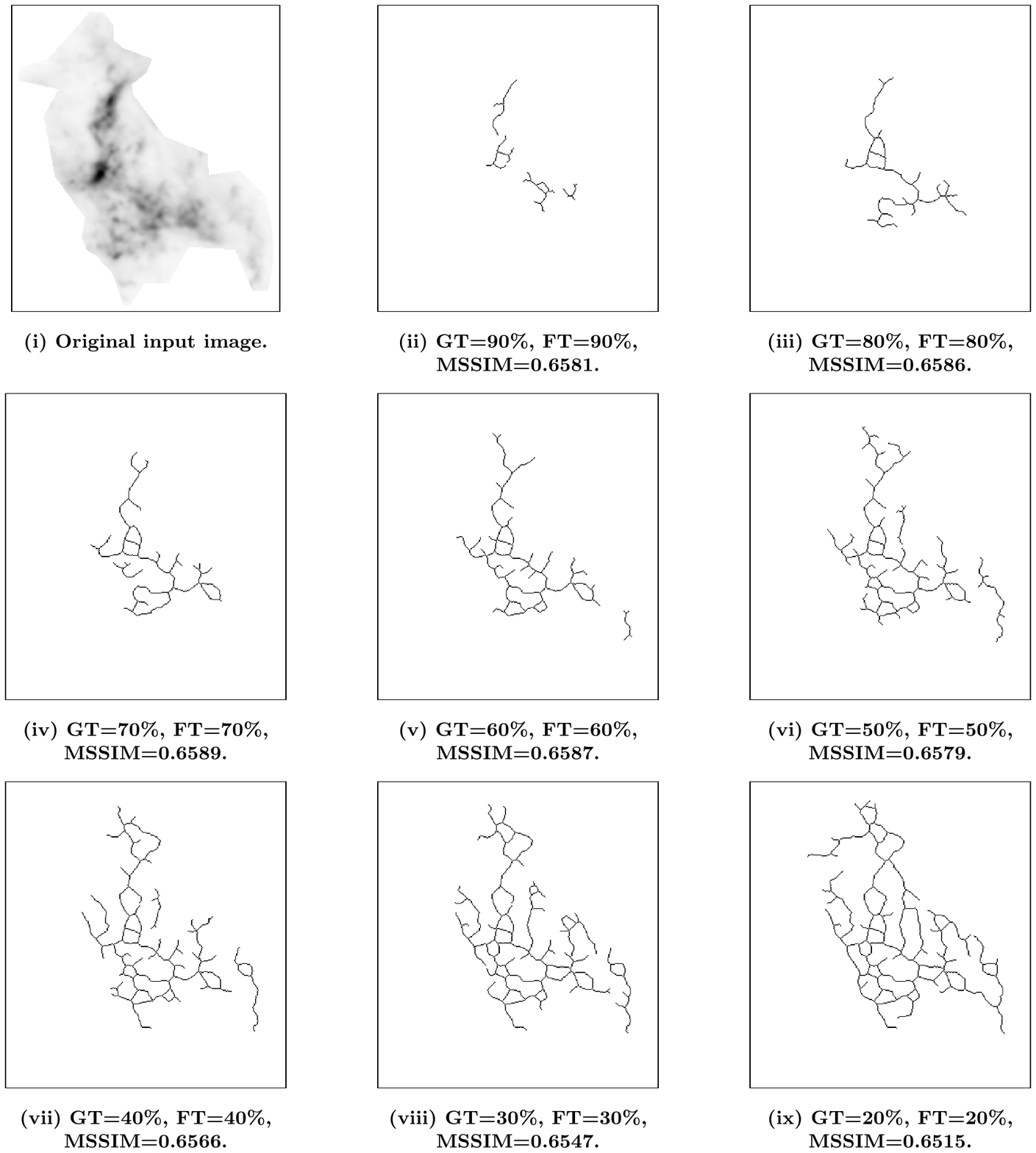}
\caption{\textbf{Different output skeletons produced by \textsc{filfinder}}. Skeletons produced by different combinations of the input parameters in a parameter study that varies the global (GT) and flattening thresholds (FT) from 0-100\% with an increment of 5\%. For the purpose of this illustration we hold the skeleton threshold at 10 pixels (0.3\,pc, corresponding to an aspect ratio of 3, given an assumed width of 0.1\,pc), and the branch threshold at 6 pixels (0.18\,pc). The input image is the dust column density map of \citet{fissel2016} for the Vela C South-ridge region. Skeletons are labelled with their corresponding input parameters, and the mean structural similarity index (MSSIM) comparing the skeleton to the original input image in panel i). \label{fig:skel_examples}}
\end{figure*}

\textsc{filfinder} is a new algorithm for filament identification \citep{koch2015}. It has four independent input parameters that influence the output: the ``skeleton threshold", the ``branch threshold", the ``global threshold", and the ``flattening threshold". These parameters respectively set the minimum length of the skeletons in pixels, minimum length of the branches in pixels, a noise threshold as a percent, and a threshold for the arctan flattening step as a percent (which removes the effect of compact, bright regions like cores from the filament mask). The output skeleton differs significantly when different input parameters are used, as illustrated in~\autoref{fig:skel_examples}. \\ 

\begin{figure*}[p]
       \centering
       
      \includegraphics[width=\textwidth, clip=true, trim=1.8cm 4.5cm 1.8cm 4cm]{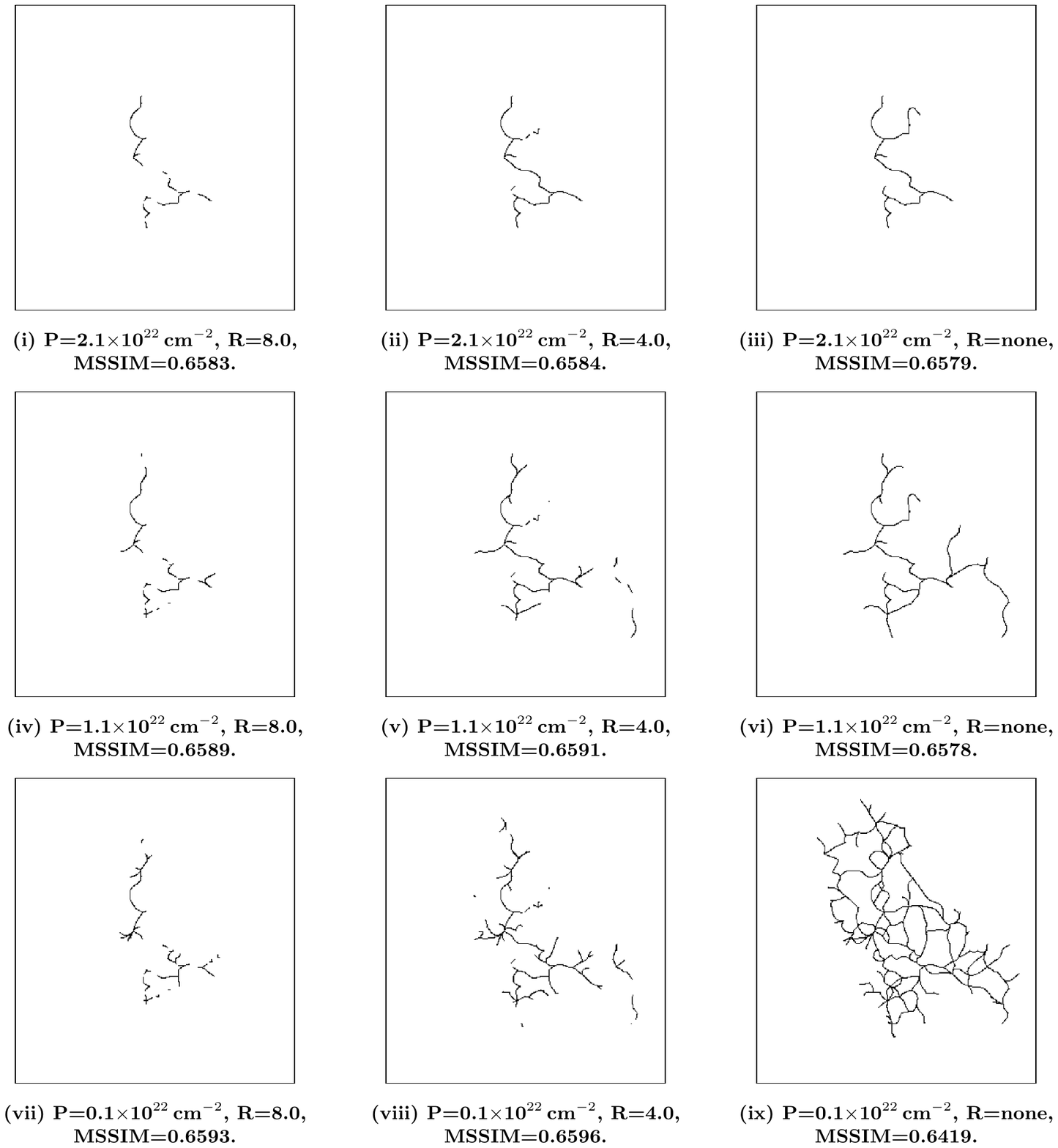}
\caption{\textbf{Different output skeletons produced by \textsc{disperse}}. Skeletons produced by different combinations of the input parameters in a parameter study that varies the persistence (P) from 0.1-4.1$\times$10$^{22}$\,cm$^{-2}$ with an increment of 0.2$\times$10$^{22}$\,cm$^{-2}$. The robusteness ratio (R) is varied from 0 (proxy for unset) to 10 with an increment of 0.5. The input image is the same as that in~\autoref{fig:skel_examples} panel i). These skeletons have not been post-processed to adjust them to fit the definition of a filament (e.g. in terms of aspect ratio). \label{fig:skel_examples_disperse}}
\end{figure*}

\textsc{disperse} is another algorithm used for filament identification \citep{sousbie2011}. With this algorithm the user sets an input image and two main input parameters called the ``persistence" and ``robustness" thresholds. \textsc{disperse} then outputs a skeleton. The persistence threshold is used to filter noise. Components of the topology of the data are represented by pairs of critical points, one negative and one positive, and these are called persistence pairs. The absolute difference of the pair is its persistence. Noise only creates or destroys topological components of persistence lower than its local value. When setting the persistence threshold the user is removing topological components with persistence lower than that threshold and is thus filtering out noise. The robustness parameter is a measure of the contrast between the filaments and the background. The robustness parameter can be set as a ratio called the robustness ratio. Again, the output skeleton differs significantly when different persistence and robustness ratio thresholds are set as shown in~\autoref{fig:skel_examples_disperse}, as would be the case for other algorithms.\\

Commonly a parameter study is performed with these algorithms to explore the effect of different input parameter combinations. These parameter studies can produce thousands of output skeletons. These may not all be good representations of the structures in the input image (e.g. the skeleton in panel ix) of~\autoref{fig:skel_examples}). Additionally some of these skeletons may be mathematically better representations than others by some predefined metric. Visually, however, some may appear to be very similar (e.g. the skeletons in panels iii) and iv) of~\autoref{fig:skel_examples}), and many may seem to be reasonable representations.\\ 


\subsection{Problems with visual selection of skeletons} \label{sec:current_skel_problem}
Identifying individual skeletons that are visually reasonable is tedious and time consuming. It is also difficult to justify why skeleton A was chosen over skeleton B when both are similar. Not only does it introduce visual bias, but the subjective nature of visual selection of skeletons means that it is often difficult to reproduce results upon repetition.  The application of the MSSIM as a mathematical goodness-of-fit measure, addresses all of these problems. \\

With the imminent influx of enormous amounts of data that astronomers face in the near future, automation of data processing is essential. The MSSIM further facilitates the automation of filament identification by eliminating the need for manual, visual selection (or rejection) in the post-processing of the skeletons. The application of the MSSIM is particularly advantageous for large datasets e.g. ALMA studies of filamentary regions.\\

\subsection{Solution: the MSSIM} \label{sec:SSIM_solution}
The similarity between two images (a `perfect' reference image and a test image) with the same dimensions can be measured using the MSSIM \citep{wang2004}. The MSSIM is a value between 1 (perfect match) and -1 (no match), providing a very simple and easy to interpret similarity measurement (Wang, private communication 2017). It is designed to be a universal index independent of the images being compared and the individual observers \citep{wang2002}. It is symmetric so the same MSSIM would be obtained if the reference and test image were switched in the comparison.\\

 The local structural similarity index (SSIM) measures the similarity between two images, $a$ and $b$, in terms of ``luminance" (mean intensity, $l(a,b)$),  ``contrast" (standard deviation of intensity, $c(a,b)$) and ``structure"\footnote{Structure is also defined as the spatial correlation and/or connectedness of pixels in the image that represent objects in the scene. In the context of the MSSIM it can also be described as any change between the two images that is not a simple shift in luminance or contrast.} (pattern stored in the image, $s(a,b)$), on a pixel-by-pixel basis. For each pixel the multiplicative combination of these three terms results in the local SSIM \citep{wang2004}:
\begin{alignat}{1}
SSIM(a, b)&=[l(a, b)]^{\alpha} \cdot [c(a, b)]^{\beta} \cdot [s(a, b)]^{\gamma} \label{eq:ssim} \\
where&: \nonumber \\
l(a, b)&= \frac{2\mu_{a}\mu_{b}+C_{1}}{\mu_{a}^{2}+\mu_{b}^{2}+C_{1}}  \\
c(a, b)&= \frac{2\sigma_{a}\sigma_{b}+C_{2}}{\sigma_{a}^{2}+\sigma_{b}^{2}+C_{2}}  \\
s(a, b)&= \frac{\sigma_{ab}+C_{3}}{\sigma_{a}\sigma_{b}+C_{3}}
\end{alignat}
where  $\mu_{a}$, $\mu_{b}$ are the local means, $\sigma_{a}$, $\sigma_{b}$ are the standard deviations and $\sigma_{ab}$ is the cross-variance for the two images. The constants $C_{1}$, $C_{2}$ and $C_{3}$, are small values to avoid the situation where the denominators would otherwise be equal to zero. In~\autoref{eq:ssim}, $\alpha=\beta=\gamma=1$ and $C_{3}=C_{2}/2$, simplifying the local SSIM to \citep{wang2004}:
\begin{equation}
SSIM(a, b)=\frac{(2\mu_{a}\mu_{b}+C_{1})(2\sigma_{ab}+C_{2})}{(\mu_{a}^{2}+\mu_{b}^{2}+C_{1})(\sigma_{a}^{2}+\sigma_{b}^{2}+C_{2})}
\end{equation}

The local SSIM values for each pixel over the images are then averaged to give the mean SSIM (MSSIM). The MSSIM is ``mostly insensitive" \citep{brooks2008} to differences between images due to changes in luminance and contrast. The luminance term is present to take care of the fact that local differences between images are less obvious in brighter regions. The contrast term accounts for local differences between images being less obvious when there is `texture'\footnote{Texture in this context can be described as rapid changes in contrast between spatially proximate pixels. The contrast term accounts for the fact that it is hard to see distortion in a highly textured image because of rapid changes in contrast, e.g. in an image of gravel it would be hard to see a distortion because of the changes in contrast between lighter rocks and their darker shadows.} in the image. According to \citet{brooks2008} the MSSIM is instead sensitive to the changes between images that ``break down natural spatial correlation" (i.e. to changes in structure), so can be used effectively to compare the structures in the original image input to the filament finding algorithm to those in the output skeleton images. \\

The MSSIM was originally used to measure image quality in television transmission, comparing the originally transmitted (reference) image to the received (degraded test) image. In this context, the output skeletons can be viewed as a very degraded version of the input image. The MSSIM can be used to compare the skeleton (degraded test) image to the input (reference) image to measure the similarity or `quality' of the skeleton with respect to the input image. The MSSIM has been used in medicine for a similar purpose, to turn endoscopic videos into still images by finding video frames that are most different to each other, reducing a lengthy video into a number of still images that form the basis of a diagnosis \citep{low2011}. In this letter we propose for the first time that the MSSIM be utilised in the post-processing of identified filament skeletons to find the skeletons most similar to the input image and thus apply it as a goodness-of-fit measure.  \\

\subsection{Using the MSSIM} \label{sec:using_SSIM}
The individual details of projects in filament identification differ. What they have in common is that many output skeletons are produced by different input parameters (where one combination produces one skeleton, and a different combination results in a different skeleton) and that these skeletons are usually saved as image arrays, generally as Flexible Image Transport System (FITS\footnote{\href{http://fits.gsfc.nasa.gov/fits\textunderscore primer.html}{http://fits.gsfc.nasa.gov/fits\textunderscore primer.html}}) files. The MSSIM can then be calculated between the original greyscale image and the black and white output skeletons using inbuilt functions within \textsc{matlab} or the \textsc{python scikit-image library} \citep{vanderwalt2014}.\\

The MSSIMs will not be large (approaching unity) as a binary black/white skeletal image is being compared to a greyscale image including more diffuse structures. In our examples in~\autoref{fig:skel_examples} and ~\autoref{fig:skel_examples_disperse} the resultant MSSIM values for the different skeletons are very close. The differences in the MSSIM values are real, representing genuine differences in the relationship between the skeleton image and input image. We are mainly interested in a method of comparison between the images that is quantitative and repeatable so even these small differences in MSSIM are sufficient. This does not negatively affect the ability of the MSSIM to discriminate between skeletons to select a quantitatively-defined `best' skeleton(s). The MSSIM effectively standardises the selection of the `best' skeleton(s) by rigorous mathematical and reproducible means.\\ 

The MSSIM is a method to compare the similarity of a skeleton to its corresponding input image. We suggest it is best utilised to find the optimum skeletons for a given input image and filament finding algorithm. It could also be used to find the best skeletal representations of the input image independently of the algorithm. However we urge caution in that approach because filament identification algorithms operate on different mathematical bases, implicitly defining filaments in different ways. They also have different input parameters and their skeletons may require different amounts of post-processing before they meet the users definition of a filament (e.g. in the removal of skeleton sections that are too short to be a filament by the users definition) and before the outputs from different algorithms are directly comparable. Additionally the MSSIM does not indicate that a skeleton and the input parameters which created it are physically sensible, nor does it indicate anything about the `goodness' or suitability of an algorithm, or input image (e.g. in terms of noise properties), for the purpose of filament identification. It only quantifies you how good a representation of the input image the corresponding skeleton is. \\

We have proposed that the MSSIM method can be used for skeleton-based algorithms, but it could also be applied for the same purpose to the outputs of non-skeleton based filament identification algorithms. These algorithms, such as the \citet{schisano2014} Hessian matrix approach, generally produce a `region of interest' (ROI) mask, in FITS file format, that defines the extent (in terms of length and width) of the detected filaments. The ROI mask is later applied to the original image to extract only the filamentary regions. Like filament skeletons, the ROI mask is a representation of the filaments in the image. To measure how good a representation of the filaments an ROI mask is, it could be compared back to the original input image using the MSSIM. Alternatively the mask could be skeletonised, and then compared. In this way the ROI mask(s) best representing the input image could be found from the pool output by the algorithm.\\

\section{Parameter optimisation with the MSSIM} \label{sec:optimisation}
\begin{figure*}[t]
       \centering
\includegraphics[width=\textwidth, clip=true, trim=1cm 6.3cm 1cm 6cm]{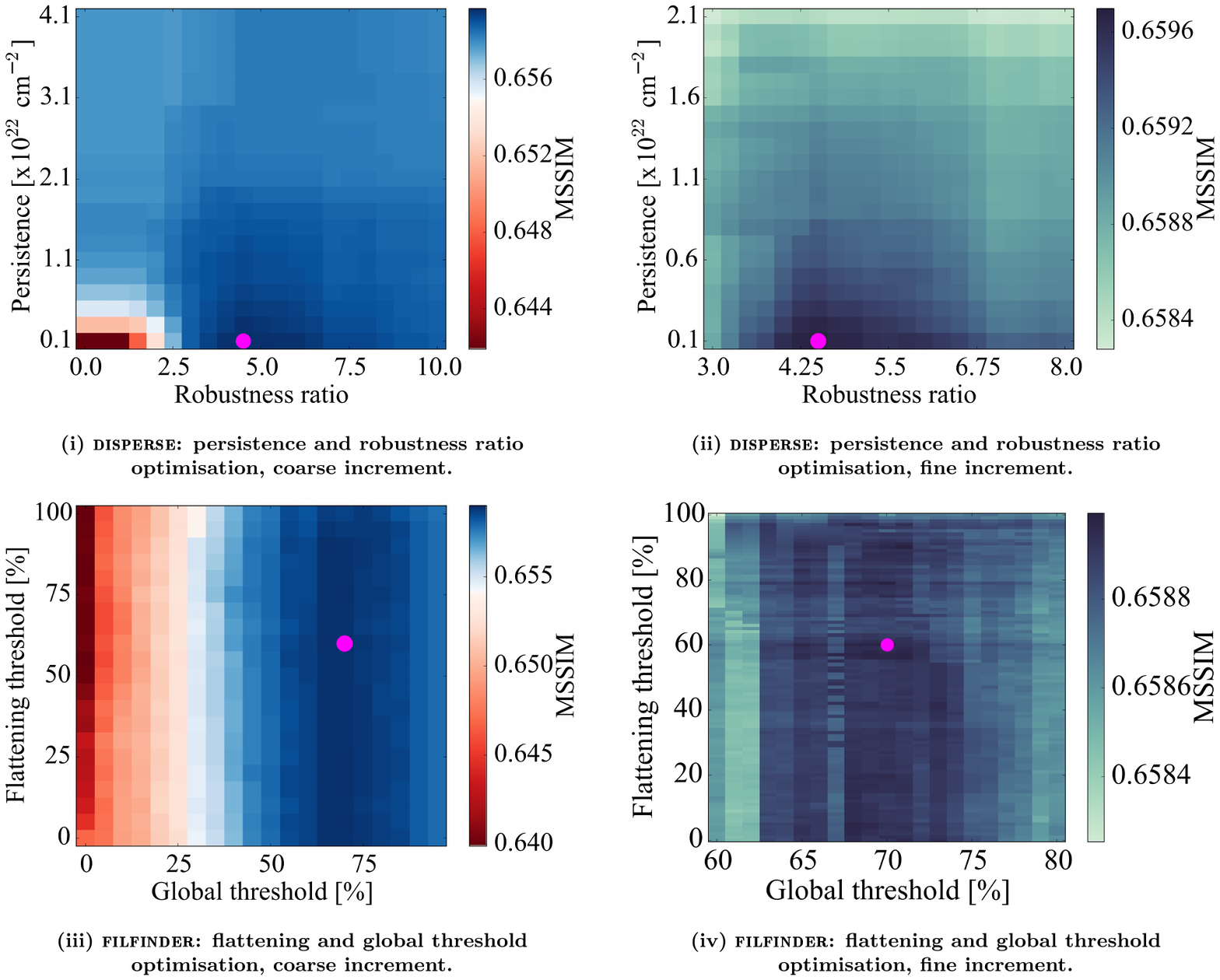}
\caption{\textbf{Optimising parameter studies using the MSSIM}. We plot as a colour map the MSSIM values of the skeletons, alongside the corresponding input parameters for the \textsc{disperse} and \textsc{filfinder} parameter studies described in the text. The `best' skeleton, with the greatest MSSIM, in each study is denoted by a magenta dot. \label{fig:param_study}}
\end{figure*}

The MSSIM makes it possible to perform systematic parameter studies exploring the impact of input parameters on skeletons. It can be used to locate the optimum range of input parameters for filament finding. An example of input parameter optimisation using the MSSIM is presented in~\autoref{fig:param_study} for both \textsc{disperse} and \textsc{filfinder}. We use the same input image as previously.\\

For \textsc{disperse} we first perform a coarsely incremented study in~\autoref{fig:param_study} panel i), exploring persistence in the range of 0.1-4.1$\times$10$^{22}$\,cm$^{-2}$ with an increment of 0.2$\times$10$^{22}$\,cm$^{-2}$. The robustness ratio is varied from 0 (proxy for unset) to 10 with an increment of 0.5. That figure shows that the highest MSSIMs are located in the persistence range of 0.1-2.1$\times$10$^{22}$\,cm$^{-2}$, and a robustness ratio range of 3-8. We then zoom in on this range in panel ii) with a more finely incremented study, with increments of 0.1$\times$10$^{22}$\,cm$^{-2}$, and 0.25 respectively. This two-stage approach of using a coarsely incremented study over a large parameter range then narrowing that range and performing a fine parameter study saves computing time compared to performing a finely incremented study over the whole parameter range. From the resulting pool of skeletons we can locate the one with the highest MSSIM, which was produced by a persistence of 0.1$\times$10$^{22}$\,cm$^{-2}$ and a robustness ratio of 4.5.\\

A similar parameter study can be performed for \textsc{filfinder}. The four parameters could be optimised in a four dimensional study, and the MSSIM used to identify the optimum skeletons from the pool produced. Alternatively the flattening and global thresholds could be optimised with the MSSIM in a two dimensional study, holding the skeleton and branch thresholds constant at values appropriate for the dataset (these are minimum cutoffs so could be reasonably set to minimum values deemed suitable). We show an example of a two stage, two dimensional \textsc{filfinder} parameter study in ~\autoref{fig:param_study}. Panel iii) of that figure shows the coarsely incremented parameter study, exploring both the global and flattening thresholds from 0-100\%, with an increment of 5\%. For the purpose of this illustration the skeleton and branch thresholds are held constant at 10 pixels (0.3\,pc, corresponding to the filament definition of an aspect ratio of 3 \citep{panopoulou2014}, given an assumed width of 0.1\,pc \citep{arzoumanian2011}) and 6 pixels (0.18\,pc) respectively. That panel shows that the highest MSSIMs occur in the global threshold range of 60-80\% so we zoom in on this range in the finely incremented parameter study shown in panel iv) with an increment of 1\%. The skeleton with the highest MSSIM in that study was produced by a global threshold of 70\% and a flattening threshold of 60\%. Regardless of the filament identification algorithm used, the MSSIM can be utilised to optimise the input parameters and identify the optimum skeleton(s) for subsequent analyses.\\

\section{Summary} \label{sec:summary}

Previously there has been no `goodness-of-fit' measure to apply to skeletons output by filament finding algorithms. Different input parameter combinations can produce thousands of output skeletons. Different algorithms also produce different skeletons. Until now the only way to select skeletons for further analysis has been visually. This comes with inherent problems, including visual bias, large time-cost and limited reproducibility. We propose the application of an existing measure used in television transmission and in medical imaging, the mean structural similarity index (MSSIM), as a `goodness-of-fit' measure for the output skeletons of filament identification algorithms. The MSSIM allows the best output skeletons for a given input image to be found for a given algorithm, or independently of the algorithm. This measure makes possible the optimisation of parameter space searches (parameter studies). It enables the user to examine the similarity of the output skeletons to the input image and immediately zoom in on the region of parameter space in which best fit solutions are located. The MSSIM for the first time presents a method by which best fit filament skeletons may be found in a completely automated and reproducible way for large volumes of data. Such a measure is essential in this era of `big data'. \\



\acknowledgements
\noindent \textbf{Acknowledgements} \\
We sincerely thank the referee for their helpful comments that improved this letter. Herschel is an ESA space observatory with science instruments provided by European-led Principal Investigator consortia and with important participation from NASA. L.M.F. and G.N. acknowledge support from NASA grant NNX13AE50G and from the Center for Interdisciplinary Exploration and Research in Astrophysics. L.M.F. was supported in part by an NSERC Postdoctoral Fellowship. LMF is a Jansky Fellow of the National Radio Astronomy Observatory (NRAO). NRAO is a facility of the National Science Foundation (NSF) operated under cooperative agreement by Associated Universities, Inc. This research made use of  \textsc{astropy} (\href{http://www.astropy.org}{http://www.astropy.org}) \citep{astropy2013}, \textsc{aplpy} (\href{http://aplpy.github.com}{http://aplpy.github.com}), Karma visualisation tools \citep{gooch1996}, \textsc{scikit-image} \citep{vanderwalt2014} and NASA's Astrophysics Data System. \\


\end{document}